\documentstyle[preprint,prl,aps]{revtex}
\newcommand{\sr}{\scriptstyle}
\begin{document}

\draft

\title{Dynamics of cylindrical domain walls in nematic
liquid crystals}

\author{Joachim Stelzer and Henryk Arod\'z}
\address{Instytut Fizyki, Uniwersytet Jagiello\'nski,
Reymonta 4, 30-059 Krak\'ow, Poland}
\date{\today}
\maketitle
\widetext

\begin{abstract}
Analytical calculations of the dynamics of a curved 
domain wall in a nematic liquid crystal are performed. 
The core of the wall is 
assumed to form a cylinder, whose axis coincides with the 
direction of an external magnetic field. 
The equation of motion for the nematic director field
is solved in a comoving coordinate frame by
applying a polynomial expansion of the tilt angle with
respect to the radial distance from the wall core. 
Starting from a cylindrical domain wall at rest as initial 
conditions, the shrinking of the cylinder and the change of the
wall width is analysed in detail. In particular, we find that 
the N\'eel wall decays faster than the Bloch wall, 
in agreement with energy considerations. 
\end{abstract}
\narrowtext

\section{INTRODUCTION}

It is a well-known fact that nematic liquid crystals reveal a large variety 
of defect structures. Such defects are very interesting from the 
theoretical point of view \cite{Chandrasekhar,Degennes1,Kleman}. 
Moreover, they are 
important for applications, {\em e.g.}, they play an
crucial r\^ole  in the switching mechanisms in modern liquid crystal
display devices (surface-stabilized ferroelectric  liquid crystal cells)
\cite{Ouchi,Yamada,Maclennan,Seitz}.
Therefore, a quantitative investigation of defect dynamics is
desirable.  

The types of defects in the nematic director field
cover zero- and one-dimensional objects (point and line
defects, respectively) that arise due to the
presence of impurities or surfaces \cite{Kleman,Nehring}. 
Two-dimensional sheets, {\em i.e.}~walls, 
are unstable in an isolated nematic. They can be stabilized,
however, by imposing an external magnetic field \cite{Helfrich}. The
situation is similar to domain walls in a ferromagnetic material
where regions of different spin orientations are separated by a thin
boundary layer \cite{Kittel}. In a nematic, the director can align 
parallel or antiparallel to the external magnetic field. Due to fact that
the director ``lacks the arrowhead'', these two situations are energetically
equivalent. It may happen, that two spatial regions of antiparallel
director orientation will form, with a domain wall between them
\cite{Helfrich}.

Domain walls in nematics due to the existence of an external field
are static in a planar geometry \cite{Helfrich}. A qualitatively new 
behaviour occurs, when the shape of the wall is curved. 
In this case the wall becomes a dynamical object,
{\em i.e.}~its core starts to move, whereas the wall width is changing.
The present paper is devoted to a quantitative study of the 
dynamics of domain walls in a
cylindrical geometry. We consider the case of director reorientation
through a Bloch and N\'eel wall \cite{Bloch,Neel}. 
From a methodological point of view, 
in our calculations we use the so-called {\em polynomial 
approximation}  -- it has turned out to be very fruitful in studies of 
defect dynamics in relativistic 
scalar field theories \cite{Arodz}. Due to the
cylindrical symmetry,  we can find explicit analytical solutions to
equations of motion obtained within that approximation.

The organization of the paper is as follows. In section II the
equation of motion for the director field is derived on the basis
of the Ericksen-Leslie theory \cite{Ericksen,Leslie} for the case of a 
cylindrical domain wall of Bloch or N\'eel type in a nematic liquid crystal. 
Section III develops the polynomial
expansion as an analytical method to solve approximately the dynamics of the 
director field. Within this approach, we obtain closed expressions
for the dynamics of the wall core and width. In section IV some
selected results together with their discussion are presented.
Section V contains an outlook on possible extensions of our
investigations.

\section{DIRECTOR EQUATIONS OF MOTION FOR CYLINDRICAL DOMAIN WALLS}

\subsection{Comoving polar coordinates}

For all calculations we restrict ourselves to the cylindrical geometry
as pointed out in Figs.\ 1 and 2. The cylinder axis is determined
by the orientation of the external magnetic field ${\bbox{H}}$
which coincides with the $z$-axis of our coordinate frame, 
{\em i.e.}~${\bbox{H}}=H_{0}\,\hat{\bbox{z}}$. ($\hat{\bbox{z}}$ 
denotes the unit vector along $z$.)  As already mentioned above, a 
domain wall occurs for parallel and antiparallel director orientation with 
respect to the external field in different spatial regions. In the case of 
axial geometry, we assume the director to be aligned along the
positive or negative $z$-axis ({\em up, down}) 
inside or outside the cylinder, respectively.
Now the director can reorient from {\em up} to {\em down} in  
either  $\hat{\bbox{\phi}}-\hat{\bbox{z}}-$plane or 
$\hat{\bbox{\rho}}-\hat{\bbox{z}}-$plane, where $(\rho,\phi,z)$ are 
cylindrical coordinates. We refer to these two
cases as {\em Bloch} (Fig.~1) and {\em N\'eel} (Fig.~2) wall, 
taking over the nomenclature from planar domain walls in 
ferromagnetics \cite{Kittel} and nematics \cite{Helfrich}.

For our calculations it will be advantageous to work within a comoving
cylindrical coordinate frame. The radial coordinate consists of the wall
radius $\rho(t)$ and the radial distance $\xi$ from the wall.
These are related to cartesian coordinates according to
\begin{equation}
x = (\rho(t)+ \xi)\,\cos\phi,\qquad
y = (\rho(t)+ \xi)\,\sin\phi.
\end{equation}
The Nabla operator, expressed in the comoving coordinates, reads
\begin{equation}
{\bbox{\nabla}} = \hat{\bbox{\xi}}\,\frac{\partial}{\partial\xi}
+ \hat{\bbox{\phi}}\,\frac{1}{\rho + \xi}\,
\frac{\partial}{\partial\phi}
+ \hat{\bbox{z}}\,\frac{\partial}{\partial z}.
\end{equation}
The ``spatial'' time derivative ${\partial\;/\partial t}$ is related
to the ``material'' ({\em i.e.}~comoving) time derivative $D\;/Dt$ 
by the chain rule between the coordinate sets $x_{i} = (x,y,z)$ and 
$\xi_{\alpha} = (\xi,\phi,z)$
(laboratory and comoving coordinates):
\begin{equation}
\frac{\partial}{\partial t} = \frac{D}{Dt} - 
\left. \frac{\partial x_{i}}{\partial t}\right|_{\xi_{\alpha}}\,
\frac{\partial \xi_{\alpha}}{\partial x_{i}}\,
\frac{\partial}{\partial \xi_{\alpha}}.
\end{equation}
The relation above is simplified for cylindrical geometry
\begin{equation}
\frac{\partial}{\partial t} = \frac{D}{Dt} - \dot{\rho}\,
\frac{\partial}{\partial\xi}.
\end{equation}
For the Bloch and N\'eel wall we can describe the 
director orientation by the {\em tilt angle field} 
$\Theta(\xi,t)$ which measures the angle between the local director
and the $z$-axis. In terms of the comoving frame, the director
is in the $\hat{\bbox{\phi}}-\hat{\bbox{z}}-$plane for the Bloch wall
and in the $\hat{\bbox{\xi}}-\hat{\bbox{z}}-$plane for the N\'eel
wall:
\begin{eqnarray} \label{dirfield}
{\bbox{n}}_{B} &=& 
\sin\Theta(\xi,t)\,\hat{\bbox{\phi}}(\phi,t)
+ \cos\Theta(\xi,t)\,\hat{\bbox{z}},\qquad \mbox{(Bloch wall)} \\
{\bbox{n}}_{N} &=& 
\sin\Theta(\xi,t)\,\hat{\bbox{\xi}}(\phi,t)
+ \cos\Theta(\xi,t)\,\hat{\bbox{z}}.\qquad \mbox{(N\'eel wall)} 
\end{eqnarray}
\subsection{Free energy density}
In the framework of the Ericksen-Leslie theory \cite{Ericksen,Leslie} 
the dynamics of the director field follows from the balance between
elastic, magnetic and viscous torques. The latter are determined by
the temporal change in the director, whereas elastic and magnetic
torques are obtained as a variational derivative from a free energy
density ${\cal F}$. Thus the equation of motion for the director
components generally reads
\begin{equation} \label{npoint}
\gamma_{1}\,\frac{\partial n_{i}}{\partial t} 
= - \frac{\delta{\cal F}}{\delta n_{i}}
\equiv \partial_{j}\,\frac{\partial{\cal F}}{\partial(\partial_{j}n_{i})}
- \frac{\partial{\cal F}}{\partial n_{i}}.
\end{equation}
In Eq.\ (\ref{npoint}) $\gamma_{1}$ is the rotational viscosity of
the liquid crystal. In fact, Eq.\ (\ref{npoint}) should contain 
an additional term $\lambda n_{i}$, with the Lagrange multiplier
$\lambda$, accounting for the constraint of ${\bbox{n}}$ being
a unit vector. In the cylindrical geometry under consideration,
however, it turns out, that this constraint can be incorporated
much more easily by starting from (\ref{npoint}) ({\em i.e.\ }
without the Lagrange multiplier). Later on we shall eliminate the
variation of the length of the director by taking appropriate
projections, as will be demonstrated below.

The free energy density ${\cal F}$ consists of an elastic and a
magnetic part (${\cal F} = 
{\cal F}_{\mathrm{elast}} + {\cal F}_{\mathrm{mag}}$). 
For the elastic free energy density we take the
Oseen-Z\"ocher-Frank expression \cite{Oseen,Zocher,Frank}, that contains
{\em splay}, {\em twist} and {\em bend} deformations of the
director field. 
\begin{equation} \label{frankenergy}
{\cal F}_{\mathrm{elast}} = \frac{\sr 1}{\sr 2}\,K_{11}\,
(\mbox{div}\,{\bbox{n}})^{2}
+ \frac{\sr 1}{\sr 2}\,K_{22}\,
({\bbox{n}}\cdot\mbox{curl}\,{\bbox{n}})^{2}
+ \frac{\sr 1}{\sr 2}\,K_{33}\,
(\bbox{n}\times\mbox{curl}\,{\bbox{n}})^{2}.
\end{equation}
In (\ref{frankenergy}) $K_{11}$, $K_{22}$ and $K_{33}$ denote
elastic constants of the nematic liquid crystal. 

The magnetic free energy density couples the director ${\bbox{n}}$
to the magnetic field ${\bbox{H}}$ via the anisotropy of the
magnetic susceptibility $\Delta\chi$ ($\mu_{0}$ means the magnetic
field constant).
\begin{equation} \label{magenergy}
{\cal F}_{\mathrm{mag}} = - \frac{\sr 1}{\sr 2}\,\mu_{0}\,\Delta\chi\,
({\bbox{n}}\cdot{\bbox{H}})^{2}.
\end{equation}
For a cylindrical Bloch or N\'eel wall we can express the free energy
density explicitly in comoving coordinates, 
\begin{eqnarray} \label{expenergy1}
{\cal F}_{B} &=& \frac{\sr 1}{\sr 2}\,K_{22}\,\left( \Theta'
+ \frac{\cos\Theta\,\sin\Theta}{\rho + \xi}\right)^{2}
+ \frac{\sr 1}{\sr 2}\,K_{33}\,\frac{\sin^{4}\Theta}{(\rho + \xi)^{2}}
\nonumber \\
& & - \frac{\sr 1}{\sr 2}\,\mu_{0}\,\Delta\chi\,H_{0}^{2}\,\cos^{2}\Theta,
\qquad \mbox{(Bloch wall)} 
\\ \label{expenergy2}
{\cal F}_{N} &=& \frac{\sr 1}{\sr 2}\,K_{11}\,\left( \Theta' \,\cos\Theta
+ \frac{\sin\Theta}{\rho + \xi}\right)^{2}
+ \frac{\sr 1}{\sr 2}\,K_{33}\,\Theta'\,^{2}\,\sin^{2}\Theta 
\nonumber \\
& & - \frac{\sr 1}{\sr 2}\,\mu_{0}\,\Delta\chi\,H_{0}^{2}\,\cos^{2}\Theta.
\qquad \mbox{(N\'eel wall)} 
\end{eqnarray}
In (\ref{expenergy1}) and (\ref{expenergy2}) $\Theta'$ stands for
the radial derivative $\partial\Theta/\partial\xi$.

\subsection{Director equation of motion}

Returning to Eq.\ (\ref{npoint}) we now have to account for the
constraint of the director normalization ($|{\bbox{n}}|=1$).
To this aim we write the left-hand side (LHS) of (\ref{npoint})
explicitly in comoving coordinates.
\begin{eqnarray} \label{LHSbloch}
\mbox{LHS}_{B} &=& - \gamma_{1}\,(\dot{\Theta} - \dot{\rho}\,\Theta')\,
(\hat{\bbox{\xi}}\times{\bbox{n}}), 
\qquad \mbox{(Bloch wall)}
\\ \label{LHSneel}
\mbox{LHS}_{N} &=& \gamma_{1}\,(\dot{\Theta} - \dot{\rho}\,\Theta')\,
(\hat{\bbox{\phi}}\times{\bbox{n}}). 
\qquad \mbox{(N\'eel wall)}
\end{eqnarray}
Here $\dot{\Theta}$ denotes the material time derivative
$D\Theta/Dt$.

For further considerations we express the right-hand side (RHS) 
of (\ref{npoint}) in cartesian coordinates. Its components read
\begin{eqnarray} \label{RHS}
\mbox{RHS}_{i} &=& K_{11}\,\partial_{i}\,(\mbox{div}\,{\bbox{n}})
- K_{22}\,(\bbox{n}\cdot\mbox{curl}\,\bbox{n})\,
(\mbox{curl}\bbox{n})_{i} 
\nonumber \\
& & - K_{22}\,\{
\mbox{curl}\,[({\bbox{n}}\cdot\mbox{curl}{\bbox{n}})\,
{\bbox{n}}]\}_{i}
\nonumber \\
& & - K_{33}\,\{ (n_{k}\,\partial_{k}n_{j})\,\partial_{i}n_{j}
- \partial_{k}[n_{k}(n_{j}\,\partial_{j}n_{i})]\} 
\nonumber \\
& & + \mu_{0}\,\Delta\chi\,({\bbox{n}\cdot{\bbox{H}}})\,H_{i}.
\end{eqnarray}

The constraint of the director normalization $|{\bbox{n}}|= 1$
means that we have to discard the (infinitesimal) variation of the length
of the director $\delta{\bbox{n}}=\epsilon({\bbox{r}})\,
\bbox{n}$. ($\epsilon(\bbox{r})$ is a small number everywhere.)
The corresponding variation of the total
free energy $F$ (which is the volume integral over the free energy 
density ${\cal F}$) reads
\begin{equation} \label{varlength}
\delta F = \int\,\mbox{d}^{3}r\,
\frac{\delta {\cal F}}{\delta {\bbox{n}}}\cdot
\delta{\bbox{n}} = \int\,\mbox{d}^{3}r\,\epsilon(\bbox{r})\,
\bbox{n}\cdot\frac{\delta{\cal F}}{\delta{\bbox{n}}} = 0. 
\end{equation}

From Eq.\ (\ref{varlength}) it is obvious that {\em the
variation of the director length is related to the
projection of the variational derivative of
the free energy density} ({\em i.e.~}of the RHS (\ref{RHS})) 
{\em onto the director.} 
Moreover, from (\ref{LHSbloch}), (\ref{LHSneel}) 
we recognize that  
the LHS, projected onto the director, yields zero. Thus we 
conclude that by discarding the projection of 
Eq.~(\ref{npoint}) onto the director
we properly take into account the director constraint!

Additionally, after some lengthy calculations we find that for the 
Bloch and N\'eel wall the projection onto $\hat{\bbox{\xi}}$ and 
$\hat{\bbox{\phi}}$, respectively, yield identically zero. 
Thus the scalar
equation of motion for the tilt angle field $\Theta(\xi,t)$ can be 
obtained by projecting (\ref{npoint}) onto the the third
linearly independent direction, which is 
$(\hat{\bbox{\xi}}\times\bbox{n})$ for the Bloch wall and
$(\hat{\bbox{\phi}}\times\bbox{n})$ for the N\'eel wall.
After performing the projections and changing for the comoving
coordinates, we obtain the equations of motion for the tilt angle field.
\begin{eqnarray} \label{Blochequ}
\gamma_{1}\,(\dot{\Theta} - \dot{\rho}\,\Theta')
&=& K_{22}\,\Theta'' + K_{22}\,\frac{\Theta'}{\rho + \xi}
\nonumber \\
& & - K_{22}\,\frac{\sin\Theta\,\cos\Theta}{(\rho + \xi)^{2}} 
\nonumber \\
& & - 2\,(K_{33} - K_{22})\,
\frac{\cos\Theta\,\sin^{3}\Theta}{(\rho + \xi)^{2}}
\nonumber \\
& & - \mu_{0}\,\Delta\chi\,H_{0}^{2}\,\sin\Theta\,\cos\Theta,
\qquad \mbox{(Bloch wall)} 
\\ \label{Neelequ}
\gamma_{1}\,(\dot{\Theta} - \dot{\rho}\,\Theta')
&=& K_{11}\,\cos\Theta\,\frac{\partial}{\partial\xi}\,
\left\{ \frac{1}{\rho + \xi}\,\frac{\partial}{\partial\xi}\,
[(\rho + \xi)\,\sin\Theta]\,\right\}
\nonumber \\
& & - K_{33}\,\sin\Theta\,\frac{\partial}{\partial\xi}\,
\left\{ \frac{1}{\rho + \xi}\,\frac{\partial}{\partial\xi}\,
[(\rho + \xi)\,\cos\Theta]\,\right\}
\nonumber \\
& & - K_{33}\,\frac{\sin\Theta\,\cos\Theta}{(\rho + \xi)^{2}}
\nonumber \\
& & - \mu_{0}\,\Delta\chi\,H_{0}^{2}\,\sin\Theta\,\cos\Theta.
\qquad \mbox{(N\'eel wall)} 
\end{eqnarray}
In agreement with the pictorial visualization of Figs.~1 and 2
the cylindrical Bloch wall contains no {\em splay} deformations,
whereas in the N\'eel wall {\em twist} deformations are absent.

\section{SOLUTION FOR THE DIRECTOR DYNAMICS}

\subsection{Polynomial expansion}

The method to construct an approximate solution of the equations of 
motion (\ref{Blochequ}),
(\ref{Neelequ}) is the {\em polynomial expansion of the field} 
which has been developed in  previous papers \cite{Arodz}.
We shall specify it here to the case of cylindrical domain walls in
nematic liquid crystals.

The key idea is to take the spatial dependence of the tilt angle field
$\Theta(\xi,t)$ as a Taylor-like expansion with respect to the
wall distance $\xi$ around its ``core'' value $\Theta(\xi=0,\,t)$.
The wall core corresponds to the radius of the cylinder 
({\em i.e.}~$\xi=0$), where the director is oriented perpendicular to the 
$z$-axis, thus $\Theta(\xi=0,\,t)=\frac{\pi}{2}\: \forall t$.
The temporal evolution of the field is governed by the expansion
coefficients. 
\begin{equation} \label{poly}
\Theta(\xi,t) = \frac{\sr \pi}{\sr 2} + a(t)\,\xi 
+ \frac{\sr 1}{\sr 2}\,b(t)\,\xi^{2}
+ \frac{\sr 1}{\sr 6}\,c(t)\,\xi^{3}.
\end{equation}
The polynomial expansion cannot be truncated at arbitrary order,
because we have to glue the bulk solution for $\Theta(\xi,t)$ smoothly
to the boundary conditions. This is not always possible. However,
it has been demonstrated for the case of cylindrical domain walls
in a relativistic field theory (where the same symmetry considerations
are valid as in our case), that the third-order
polynomial expansion (\ref{poly}) guarantees the compatibility of the bulk
solution with the boundary conditions \cite{Arodz}. Of course, for higher 
order of the Taylor expansion the accuracy of the calculations
will increase. 

From (\ref{poly}) we can easily obtain the spatial and temporal
derivatives of the tilt angle field $\Theta(\xi,t)$. The trigonometric
and fractional expressions that occur in the equations of motion
(\ref{Blochequ}), (\ref{Neelequ}) are expanded as well up to third
order in the wall distance $\xi$. All these expansions are then
inserted into (\ref{Blochequ}) and (\ref{Neelequ}). By comparison
of the coefficients of subsequent orders in $\xi$ we obtain 
a set of ordinary differential equations for the expansion
coefficients $a(t)$, $b(t)$ and $c(t)$. For the third-order
polynomial expansion (\ref{poly}) such a comparison yields meaningful
results for zeroth and first order in $\xi$. Thus from our
calculations we obtain two equations that read as follows:
\begin{eqnarray} \label{eq1}
\gamma_{1}\,\dot{\rho}\,a &=& - K_{22}\,b - K_{22}\,\frac{a}{\rho},
\qquad \mbox{(Bloch wall)}
\\ 
\gamma_{1}\,(\dot{a} - \dot{\rho}\,b) &=&
K_{22}\,c + K_{22}\,\frac{b}{\rho} + 2\,(K_{33} - K_{22})\,
\frac{a}{\rho^{2}}
\nonumber \\ 
& & + \mu_{0}\,\Delta\chi\,H_{0}^{2}\,a,
\qquad \mbox{(Bloch wall)}
\\
\gamma_{1}\,\dot{\rho}\,a &=& - K_{33}\,b - K_{33}\,\frac{a}{\rho},
\qquad \mbox{(N\'eel wall)}
\\ \label{eq2}
\gamma_{1}\,(\dot{a} - \dot{\rho}\,b) &=&
K_{33}\,c + K_{33}\,\frac{b}{\rho} 
\nonumber \\
& & - (K_{33} - K_{11})\,\frac{a}{\rho^{2}}
- (K_{33} - K_{11})\,a^{3}
\nonumber \\
& & + \mu_{0}\,\Delta\chi\,H_{0}^{2}\,a.
\qquad \mbox{(N\'eel wall)}
\end{eqnarray}

\subsection{Boundary conditions}

Next we have to incorporate the boundary conditions. These are
determined by the director alignment parallel and antiparallel
to the external magnetic field inside and outside the wall,
respectively. The (instantaneous)
wall thickness is given as $\xi_{0}(t)+\xi_{1}(t)$, where $\xi_{0}(t)$ 
measures
the outward distance from the core to the outer edge of the wall, whereas
$\xi_{1}(t)$ means the core-to-edge distance towards the cylinder
axis. Therefore the boundary conditions read
\begin{eqnarray} \label{bound1}
\Theta(\xi_{0},t) = \pi, & \qquad & \Theta(-\xi_{1},t) = 0,
\\ \label{bound2}
\Theta'(\xi_{0},t) = 0, & \qquad & \Theta'(-\xi_{1},t) = 0.
\end{eqnarray}
Eqs.~(\ref{bound1}) determine the director orientation at the edge
of the wall, and Eqs.~(\ref{bound2}) are the conditions for the 
smoothness of the solution.

We can immediately express the boundary conditions in terms of
the polynomial expansion for the tilt angle field (\ref{poly}):
\begin{eqnarray}
a\,\xi_{0} + \frac{\sr 1}{\sr 2}\,b\,\xi_{0}^{2}
+ \frac{\sr 1}{\sr 6}\,c\,\xi_{0}^{3} &=& \frac{\sr \pi}{\sr 2}, 
\\
a + b\,\xi_{0} + \frac{\sr 1}{\sr 2}\,c\,\xi_{0}^{2} &=& 0,
\\
- a\,\xi_{1} + \frac{\sr 1}{\sr 2}\,b\,\xi_{1}^{2}
- \frac{\sr 1}{\sr 6}\,c\,\xi_{1}^{3} &=& - \frac{\sr \pi}{\sr 2},
\\
a - b\,\xi_{1} + \frac{\sr 1}{\sr 2}\,c\,\xi_{1}^{2} &=& 0.
\end{eqnarray}
The equations above form a set of inhomogeneous linear equations for the 
expansion coefficients $a(t)$, $b(t)$, $c(t)$.
The solubility conditions yield relations between the expansion
coefficients  and the wall partial widths
$\xi_{0}(t)$ and $\xi_{1}(t)$.
\begin{eqnarray}
\xi_{0} = \xi_{1}, &\qquad & a = \frac{3\pi}{4\xi_{0}}, 
\\
b = 0, & \qquad & c = - \frac{\sr 3\pi}{\sr 2\xi_{0}^{3}} 
= - \frac{\sr 32}{\sr 9\pi^{2}}\,a^{3}.
\end{eqnarray}
We note that the partial widths of the wall have turned 
out identical, and 
that there is no quadratic term in the polynomial expansion.

These relations must hold also in the bulk equations
(\ref{eq1})--(\ref{eq2}) in order to fulfill the boundary 
conditions. Introducing them, (\ref{eq1})--(\ref{eq2}) 
become simplified.
\begin{eqnarray} \label{radius1}
\gamma_{1}\,\dot{\rho} &=& - \frac{K_{22}}{\rho},
\qquad \mbox{(Bloch wall)}
\\ \label{width1}
\gamma_{1}\,\dot{a} &=& - \frac{\sr 32}{\sr 9\pi^{2}}\,K_{22}\,a^{3}
+ 2\,(K_{33} - K_{22})\,\frac{a}{\rho^{2}}
\nonumber \\
& & + \mu_{0}\,\Delta\chi\,H_{0}^{2}\,a, \qquad \mbox{(Bloch wall)}
\\ \label{radius2}
\gamma_{1}\,\dot{\rho} &=& - \frac{K_{33}}{\rho},
\qquad \mbox{(N\'eel wall)}
\\ \label{width2}
\gamma_{1}\,\dot{a} &=& 
- (\frac{\sr 32}{\sr 9\pi^{2}}\,K_{33} + K_{33} - K_{11})\,a^{3}
- (K_{33} - K_{11})\,\frac{a}{\rho^{2}}
\nonumber \\
& & + \mu_{0}\,\Delta\chi\,H_{0}^{2}\,a. \qquad \mbox{(N\'eel wall)}
\end{eqnarray}
Equations (\ref{radius1}) and (\ref{radius2}) describe the time evolution 
of the wall core, {\em i.e.}~the cylinder radius, whereas
(\ref{width1}) and (\ref{width2}) govern the dynamics of the
wall half-width, for the Bloch and N\'eel wall, respectively.

\subsection{Solution for the wall dynamics}

The coupled equations of motion (\ref{radius1})--(\ref{width2})
can be solved analytically. First we solve (\ref{radius1}) and
(\ref{radius2}) for the dynamics of the wall core. It obeys
a square root law:
\begin{eqnarray} \label{radbloch}
\rho(t) &=& = \sqrt{\rho_{0}^{2} - 2\,\frac{K_{22}}{\gamma_{1}}\,t},
\qquad \mbox{(Bloch wall)}
\\ \label{radneel}
\rho(t) &=& = \sqrt{\rho_{0}^{2} - 2\,\frac{K_{33}}{\gamma_{1}}\,t}.
\qquad \mbox{(N\'eel wall)}
\end{eqnarray}

Qualitatively, starting from a wall radius $\rho_{0}$, the cylindrical
domain wall will shrink. This behaviour is similar to what has been
found for relativistic field theories \cite{Arodz}.
For usual nematics $K_{22}$ is the smallest
of the elastic constants. Therefore the  
(\ref{radbloch}) and (\ref{radneel}) 
indicate that the decay time of the Bloch wall
is larger than that of the N\'eel wall.

Now we can use the solutions (\ref{radbloch}), (\ref{radneel}) to
eliminate the $\rho$-dependence from the equations of motion
for the wall width (\ref{width1}), (\ref{width2}). The latter then read
\begin{equation} \label{Bernoulli}
\dot{a} = A(t)\,a - B\,a^{3},
\end{equation}
where the abbreviations stand for
\begin{eqnarray}
A(t) &=& \frac{2\,(K_{33} - K_{22})}{\gamma_{1}\,\rho_{0}^{2} -
  2\,K_{22}\,t} + \frac{\mu_{0}\,\Delta\chi}{\gamma_{1}}\,H_{0}^{2},
\qquad \mbox{(Bloch wall)}
\\
B &=& \frac{\sr 32}{\sr 9\pi^{2}}\,\frac{K_{22}}{\gamma_{1}},
\qquad \mbox{(Bloch wall)}
\\
A(t) &=& - \frac{(K_{33} - K_{11})}{\gamma_{1}\,\rho_{0}^{2} -
  2\,K_{33}\,t} + \frac{\mu_{0}\,\Delta\chi}{\gamma_{1}}\,H_{0}^{2},
\qquad \mbox{(N\'eel wall)}
\\
B &=& \frac{\sr 32}{\sr 9\pi^{2}}\,\frac{K_{33}}{\gamma_{1}}
+ \frac{K_{33} - K_{11}}{\gamma_{1}}.
\qquad \mbox{(N\'eel wall)}
\end{eqnarray}
Eq.\ (\ref{Bernoulli}) is an ordinary differential equation of 
Bernoulli type. By the substitution $a =
1/ \sqrt{\tilde{a}}$ it can be transformed into an inhomogeneous linear
differential equation
\begin{equation}
\dot{\tilde{a}} + 2\,A(t)\,\tilde{a} = 2\,B,
\end{equation}
whose general solution is obtained from a two-fold integration:
\begin{equation}
\tilde{a}(t) = \frac{1}{M(t)}\,[2\,B\,N(t) + \tilde{a}_{0}],
\end{equation}
\begin{equation}
M(t) = \exp(\int_{0}^{t}\,\mathrm{d}t'\,2\,A(t')), 
\end{equation}
\begin{equation}
N(t) = \int_{0}^{t}\,\mbox{d}t'\,M(t').
\end{equation}
$\tilde{a}_{0}$ is the initial value for the auxiliary variable
$\tilde{a}(t)$. This variable is related closely to the wall half-width
$\xi_{0}(t)$ (Eq.(51) below). 

For our case of the cylindrical domain walls all integrals can be
solved analytically. 
\begin{eqnarray} \label{M}
M(t) &=& \mbox{e}^{\alpha\,t}\,(1 - \beta\,t)^{\delta},
\\ \label{N}
N(t) &=&
\frac{\mbox{e}^{\alpha/\beta}}{\beta}\,
\left( \frac{\beta}{\alpha}\right)^{\delta  + 1}\,
\left[ \gamma(\delta + 1, \frac{\alpha}{\beta})
- \gamma(\delta + 1, \frac{\alpha}{\beta} - \alpha\,t)\right], 
\end{eqnarray}
where further abbreviations are introduced:
\begin{eqnarray} \label{abbr1}
& \alpha = 
\frac{\displaystyle 
2\,\mu_{0}\,\Delta\chi}{\displaystyle \gamma_{1}}\,H_{0}^{2},\quad
\beta = \frac{\displaystyle 
2\,K_{22}}{\displaystyle \gamma_{1}\,\rho_{0}^{2}},\quad
\delta = - 2\,\frac{\displaystyle K_{33}}{\displaystyle K_{22}} + 2, &
\qquad \mbox{(Bloch wall)}
\\
& & \nonumber \\ \label{abbr2}
& \alpha = 
\frac{\displaystyle 
2\,\mu_{0}\,\Delta\chi}{\displaystyle \gamma_{1}}\,H_{0}^{2},\quad
\beta = \frac{\displaystyle 
2\,K_{33}}{\displaystyle \gamma_{1}\,\rho_{0}^{2}},\quad
\delta = 1 - \frac{\displaystyle K_{11}}{\displaystyle K_{33}}. & 
\qquad \mbox{(N\'eel wall)}
\end{eqnarray}
In (\ref{N}) $\gamma(\nu, x)$ stands for the incomplete Gamma
function, whose integral representation is
\begin{equation}
\gamma(\nu, x) = \int_{0}^{x}\,\mbox{d}t\,\mbox{e}^{-t}\,
t^{\nu - 1}.
\end{equation}
From the solution for the auxiliary variable $\tilde{a}(t)$ the
time evolution of the wall half-width $\xi_{0}(t)$ is obtained by simple
back substitution,
\begin{equation} 
\xi_{0}(t) = \frac{3\pi}{4a(t)} = \frac{3\pi}{4}\,\sqrt{\tilde{a}(t)}.
\end{equation}

\section{QUANTITATIVE RESULTS}

The analytical results for the dynamics of cylindrical domain walls
in nematics that have been obtained in the previous section
can be specified to real materials. For further considerations we choose the 
parameters that enter the solutions according to the nematic phase of PAA 
($p$-azoxyanisole) at 120$^{\circ}$C 
\cite{Chandrasekhar,Degennes1} (Table I). 

\begin{center}
\begin{tabular}{|c|c|}
\hline
splay elastic constant $K_{11}$ & $7.0\cdot 10^{-12}\mbox{ N}$ \\
twist elastic constant $K_{22}$ & $4.3\cdot 10^{-12}\mbox{ N}$ \\
bend elastic constant $K_{33}$ & $1.7\cdot 10^{-11}\mbox{ N}$ \\
rotational viscosity $\gamma_{1}$ & $6.7\cdot 10^{-3}\,\mbox{ Nm/s}$ \\
magnetic anisotropy $\mu_{0}\,\Delta\chi$ & $1.21\cdot 10^{-7}$ \\
\hline
\end{tabular}

\vspace{0.5cm}

TABLE I: Material constants of PAA at 120$^{\circ}$C.
\end{center}
The magnetic field strength $H_{0}$ has been chosen 500 Oersted, 
according to a magnetic flux density $B_{0}\equiv \mu_{0}\,H_{0} =0.05$ T.

The initial configuration is a cylindrical domain wall at rest. 
At zero time the cylinder radius is by two or three orders of magnitude larger
than the wall half-width: $\rho_{0}\equiv \rho(t=0)= 0.1\,$mm,
with
$\tilde{\xi}_{0}\equiv\xi_{0}(t=0) = 1\,\mu$m or
$0.1\,\mu$m.

Fig.~3 shows the shrinking of the wall according to the square-root
law (\ref{radbloch}), (\ref{radneel}). The actual decay time is the
moment when the cylinder radius touches the edge of the wall.
However, from the temporal evolution of the wall width (see below) it is
obvious that the decay time is only slightly overestimated when taking
the time for which the cylinder has shrinked to zero.
\begin{eqnarray} \label{decay1}
\tau_{\mathrm{decay}} &=& \frac{\gamma_{1}\,\rho_{0}^{2}}{2\,K_{22}}, 
\qquad \mbox{(Bloch wall)} 
\\ 
\label{decay2}
\tau_{\mathrm{decay}} &=& \frac{\gamma_{1}\,\rho_{0}^{2}}{2\,K_{33}}. 
\qquad \mbox{(N\'eel wall)} 
\end{eqnarray}
The main feature of the Bloch wall are twist
deformations, whereas the N\'eel wall rather consists of bend
deformations. The decay times (\ref{decay1}), (\ref{decay2}) 
depend on the elastic constants
$K_{33}$ and $K_{22}$: the N\'eel wall decays about four times
faster than the Bloch wall.
 
Interestingly, the dynamics of the wall width reveals two
separated processes on different time scales. First 
a rapid change occurs from the initial half-width
to a metastable state that does not depend on the initial
condition (Figs.~4 and 5). The metastable half-width 
can be related to the magnetic coherence
length $\xi_{\mathrm{mag}}$. For planar geometry the director would
reorient in an external magnetic field by a {\em twist} deformation
on a length scale \cite{Degennes2} of
\begin{equation} \label{coherence}
\xi_{\mathrm{mag}} = \sqrt{\frac{K_{22}}{\mu_0\,\Delta\chi}}\,\frac{1}{H_0}
\end{equation}
Introducing the parameters given at the beginning of this section,
we obtain $\xi_{\mathrm{mag}}= 0.15\,\mu$m for the half-width 
of a
planar Bloch wall. However, according to Figs.~4 and 5
the metastable value for
the half-width of the cylindrical Bloch wall is  
$\xi_{\mathrm{Bloch}}= 0.21\,\mu$m.
The discrepancy is due to the curvature that gives rise to additional
{\em bend} deformations. Assuming that a similar law as 
Eq.~(\ref{coherence}) also holds for the curved wall, we can extract
an effective elastic constant 
$K_{\mathrm{Bloch}} = 8.4\cdot 10^{-12}$ N. 
For the curved N\'eel wall the coherence length (Figs.~4
and 5) is
$\xi_{\mathrm{Neel}} =  0.68\,\mu$m, corresponding to an
effective elastic constant of 
$K_{\mathrm{Neel}} = 8.85\cdot 10^{-11}$ N. This means that the
elastic energy content of cylindrical domain walls is about one 
order of magnitude larger for the N\'eel geometry as compared to  
the Bloch-like reorientation.

As already stated above the effective coherence length
$\xi_{\mathrm{Bloch}}$ or $\xi_{\mathrm{Neel}}$ denotes
a metastable state.  After some seconds, 
when the cylinder has shrinked so that its radius $\rho(t)$
is of the order $\xi_{0}(t)$, our solution for the
wall half-width breaks down. 
Formally, our expressions for the Bloch wall
reveal an implosion (Fig.~6), whereas for the N\'eel wall we obtain an 
explosion (Fig.~7) of the wall width. This is due to the opposite
sign of the quantity $\delta$ in (\ref{M}) and (\ref{N}) and thus
depending on the relative magnitude of the elastic constants
({\em cf.}~(\ref{abbr1}), (\ref{abbr2})). Actually, one should remember that
for wall half-widths larger than the cylinder radius the boundary conditions
(\ref{bound2}) are not correct, and then our solutions loose 
their physical meaning. Thus, within the framework of the polynomial
expansion we are able to study the decay of the cylindrical 
domain wall until its radius becomes approximately equal to its half-width.

\section{REMARKS}

1. Our solutions for $\rho(t)$ and $\xi_{0}(t)$ give a rather detailed 
description of the time evolution of cylindrical domain walls. First of all, 
it would be interesting to compare our theoretical predictions with 
experimental results. Furthermore, because our expressions explicitly show how 
the dynamics of the domain wall depends on the material constants 
$K_{11}$, $K_{22}$, $K_{33}$, $\mu_0$ and $\gamma_1$, 
they could be useful for the determination
of these constants by observing the shrinking of the cylindrical domain
walls.

2. It is possible to generalize our calculations to domain walls of a
more general shapes, {\em e.g.}~we could allow for a modulation along the $z$-axis or for
non-circular sections by the $x$-$y$-plane. 
For the relativistic scalar field theory this has been 
performed in the second of the references \cite{Arodz}. The calculations in the 
general case the are much more cumbersome than for strictly
cylindrical symmetry, 
while the main ideas of the polynomial approximation remain unchanged.
For this reason we have chosen to restrict our presentation to the 
cylindrical geometry.

\acknowledgments

As a Feodor-Lynen fellow J.~S.~gratefully
acknowledges his individual grant from the Alexander 
von Humboldt-Stiftung.

%
% 1
%
\begin{figure}
\caption{Geometry and coordinates for a cylindrical Bloch wall.
Projection onto a plane perpendicular to the cylinder axis.}
\end{figure}
%
% 2
%
\begin{figure}
\caption{Geometry and coordinates for a cylindrical N\'eel wall.
Projection onto a plane perpendicular to the cylinder axis.}
\end{figure}
%
% 3
%
\begin{figure}
\caption{Temporal evolution of the radius of a cylindrical 
domain wall in PAA at 120$^{\circ}$C. 
Initial wall radius $\rho(t=0) = 0.1\,$mm.
Solid: Bloch wall, dashed: N\'eel wall.}
\end{figure}
%
% 4
%
\begin{figure}
\caption{Temporal evolution of the half-width of a cylindrical 
domain wall in PAA at 120$^{\circ}$C. 
Initial wall half-width $\rho(t=0) = 1\,\mu$m.
Solid: Bloch wall, dashed: N\'eel wall.}
\end{figure}
%
% 5
%
\begin{figure}
\caption{Temporal evolution of the half-width of a cylindrical 
domain wall in PAA at 120$^{\circ}$C. 
Initial wall half-width $\rho(t=0) = 0.1\,\mu$m.
Solid: Bloch wall, dashed: N\'eel wall.}
\end{figure}
%
% 6
%
\begin{figure}
\caption{Catastrophic behaviour of the width of a cylindrical 
Bloch wall in PAA at 120$^{\circ}$C.}
\end{figure}
%
% 7
%
\begin{figure}
\caption{Catastrophic behaviour of the width of a cylindrical 
N\'eel wall in PAA at 120$^{\circ}$C.}
\end{figure}

\end{document}